# Mapping General System Characteristics to Non-Functional Requirements

A.Keshav Bharadwaj , T.R. Gopalakrishnan Nair
*Research and Industry Incubation Center*
*Dayananda Sagar Institutions, DSCE*
*Bangalore – 560078, INDIA*

***Abstract*-The Function point analysis (FPA) method is the preferred scheme of estimation for project managers to determine the size, effort, schedule, resource loading and other such parameters. The FPA method by International Function Point Users Group (IFPUG) has captured the critical implementation features of an application through fourteen general system characteristics. However, Non-functional requirements (NFRs) such as functionality, reliability, efficiency, usability, maintainability, portability, etc. have not been included in the FPA estimation method. This paper discusses some of the NFRs and tries to determine a degree of influence for each of them. An attempt to factor the NFRs into estimation has been made. This approach needs to be validated with data collection and analysis.**

## 1. Introduction

The Function point (FP) technique was defined as an alternative to the System Lines of Code (SLOC) method of estimation. Herein ***size*** was defined as a function of Inputs, Outputs, Inquiries, Internal files and External Interfaces. In this technique the General System Characteristics (GSC) were also introduced.

FPs are a measure of the functional size of Information systems. Here, one measures the functionality that the user requests and receives, independent of the technology used for implementation. The 14 General System Characteristics (GSCs) rate the general functionality of the application giving a Value Adjustment Factor (VAF) which is used to fine tune the Function point count.

FP counting is preferred to the other methods (SLOC method, Number of Programs) of estimating the size of work product since this is not influenced by bad design or bad code. Also, for GUI based programs, SLOC may not make any sense.

This paper is organized as follows. Section 2 surveys the two broad subsections of software applications. Section 3 surveys typical definitions for the terms 'functional requirement' and 'non-functional requirement'. In section 4 the mapping of GSCs to NFRs are discussed while in section 5 the degrees of influence are discussed. The mapping of NFRs are discussed in section 6. The extension of mapping of other NFRs are discussed in section 7. The paper ends with section 8, conclusion.

## 2. Application

There are two broad subsections of a software application, i.e., functional and non-functional requirements. Functional requirements (FRs) capture the intended behavior of the system, in terms of the services or tasks the system is required to perform, while non-functional requirements (NFRs) are requirements that impose restrictions on the product being developed. [8]. NFRs define the system properties and specify the behavioral pattern under various operating conditions. The various estimation methods help in sizing the application based on the functional requirements. However most of these methods have overlooked the influence of non-functional requirements. Although the term 'non-functional requirement' has been in vogue for more than 20 years there is still no consensus in the requirements among the engineering community regarding NFRs and how we should document and validate them. On the other hand, there is consensus that NFRs are important and can be critical for the success of a project**.** In practice it has been seen that neglect of the influence of NFRs may result in derailing of the project.

It so happens that NFRs are much more important compared to FRs. Say, if some functionality is left out it can always be supplemented by manual means but if the response time required is low and due to some mistake in the design, response time becomes very high then it will lead to the application becoming practically inoperable**.**

## 3. Defining the Terms

There is a rather broad consensus about how to define the term FRs: Martin Glinz in his article [5] quotes several authors on the definition of the terms FRs and NFRs. According to him, in these definitions the emphasis is either on *functions* or *behavior*. Attempting a synthesis, Wiegers defines FRs as "A statement of a piece of required functionality or a behavior that a system will exhibit under specific conditions."[10]; while Jacobson, Rumbaugh and Booch define it as "A requirement that specifies an action that a system must be able to perform, without considering physical constraints; a requirement that specifies input/output behavior of a system."[3].

NFRs according to Davis are defined as “the required overall attributes of the system; including portability, reliability, efficiency, human engineering, testability, understandability and modifiability.”[1]. However Kotonya and Sommerville define NFRs as “requirements which are not specifically concerned with the functionality of a system. They place restrictions on the product being developed and the development process, and they specify external constraints that the product must meet.”[4] Mylopoulos, Chung and Nixon’s definition is “……global requirements on its development or operational cost, performance, reliability, maintainability, portability, robustness, and the like, (….) There is not a formal definition or a complete list of nonfunctional requirements.” [6]

Traditionally, software teams address software quality requirements, that is, NFRs, using product-centric [6] methods. These methods are curative [2] and focus on gathering metrics and testing to examine a product after construction to determine whether it is within certain quality constraints. Another approach to addressing NFRs is called the process-oriented approach [2,3,8,13]. In the preventive approach of Mylopoulos, Chung and Nixon [6], the goal is to prevent problems with quality from being injected into the product during the requirements or design phases.

## 4. Mapping of GSCs to NFRs

The application characteristics such as performance, security, usability, etc also called as the NFRs are linked to the GSCs. This mapping of NFRs to GSCs has been adopted from Parthasarathy [7]. This paper attempts to frame a weighted measures table for some which can then be used to temper estimates made to improve accuracy.

## 5. Degree of Influence

One thing that should be noted is that there has been a sea of change in the way databases, programming languages, hardware platforms and operating systems have evolved ever since these concepts have come into use. As the days go by, more and more efficient systems are evolving which are much faster and efficient than their ancestors. This in turn leads us to conclude that the Degree of Influence (DI) parameter for many of the 14 GSCs no longer hold true in today’s state of the art systems. Now for example, take into account GSC1: *Data Communications* and GSC2: *Distributed Data Processing*. Previously applications were executed on stand-alone or locally connected machines. Processing data across locations required special coding techniques and good infrastructure. Nowadays, with the coming of internet most applications are executed across locations and distributed data processing is the order of the day.

## 6. Existing Mapping of some NFRs

The key NFRs that can be attributed to an application and their mapping to respective GSCs are derived as follows from [7].

| | |
|---|---|
| 1) Reliability | Operation Ease |
| 2) Response Time | No mapping given |
| 3) Performance | Performance, Online Update, Online Date Entry |
| 4) Security | No mapping given |
| 5) Availability | No mapping given |
| 6) Scalability | Transaction rate |
| 7) Capacity | No mapping given |

### 6.1. GSC-12 Operation Ease to Reliability

This GSC reflects on the fact as to how automated the system is. An application or the software system once installed and configured on a given platform should require no manual intervention, except for starting and shutting down. The system should be able to maintain a specified level of performance in case of software faults. It should also be able to re-establish its level of performance and to recover all the data directly affected in case of a failure in the minimum time and effort. This GSC is mapped on to the *reliability* NFR. It may be defined as “a system which is capable of re-establishing its level of performance and recovering the data directly affected in case of a failure and on the time and effort needed for it.”[7] The design criteria for reliability can be defined as self-containedness- the system should have all the features necessary for all its operations including recovering it by itself; completeness- it should be complete in itself and not dependent on anything else; robustness/integrity- it should not easily breakdown; error tolerance- it should be able to tolerate errors and rectify them and continue in its operation. There are “numerous metrics for determining reliability: mean time to failure, defect reports and counts, resource consumption, stability, uptime percentage and even customer perception.” [9].

### 6.2. GSC-3 Performance, GSC-8 Online Update and GSC-6 Online Data Entry to Performance

Real time systems have strict performance parameters like performing at the same level even during peak user times, producing high throughput, serving a huge user base, etc. The DI varies from no special performance requirements to response time being critical during all business hours and till performance analysis tools being used in the design. The performance NFR can be mapped on to this GSC partially. System should meet the desired performance expectations (response, processing time, throughput rates).” [7]. Also if online update has to take place then the performance expectations to be met are very high – fast response, low processing time and high throughput rates.

The performance NFR is also based on the Online Data Entry requirements of an application. The present day trend is to have interactive and real-time data entry. The GUI development requires a lot of effort as help has to be provided, validation to be implemented, reference information for faster data entry operations, etc. Performance when related to this GSC can be defined as "attributes of software that bear on response and processing times and on throughput rates in performing its function."[7].

### 6.3. GSC-5 Transaction Rate to Scalability

In many business applications the transaction rate increases to high peak levels once in a day or once in a week with the requirement remaining so that there has to be no dramatic increase in transaction time. This issue has to be looked into in the design, development and/or installation phases of a project. This GSC is mapped on to the *scalability* NFR. The term scalability implies "the ability to scale up to peak transaction loads." [7]. In order to achieve this the application has to be designed in such a way so that it should cater to the highest possible figures thus wasting resources when the transaction rate is low. The architecture should be designed in a multi-layered manner in complex algorithm based applications to scale up to peak transaction rates. In today's systems, this GSC does not contribute much to the DI as present day hardware and operating systems provide built-in features such as high bandwidth network, high speed storage disks with high-speed disk access timings and CPUs with high MHZ processing speed which when combined leads to built in high transaction rates.

## 7. Extension of Mapping of other NFRs

Now in this section mapping of those NFRs not already mapped will be done. Then weighted measures tables for these newly mapped NFRs will be developed. These weighted measures can then be used to temper projects estimates and improve their accuracy. Now four NFRs will be mapped to relevant GSCs and the Degree of Influence for each will be detailed. The mapping is as follows

| | |
|---|---|
| 1) Response Time | Data Communications, Distributed data processing, Performance |
| 2) Security | Multiple sites, Online Update |
| 3) Availability | Online data entry, operation ease, |
| 4) Capacity | Transaction rate, Multiple sites. |

Since each NFR can be mapped to more that one GSC it is necessary that for each NFR the Degree of Influence and level of complexity mapping be also done. This can be done as follows.

### 7.1. Response Time

Response time can be mapped to three different GSCs namely Data Communications, Distributed data processing and Performance. The DI mapping for response time can be done as shown in the table below.

| DI | Guidelines |
|---|---|
| 0 | Batch processing or a stand alone PC with no specific performance specifications |
| 1 | Batch application which produces data for use by other components of the system with certain performance and design criteria specified but not required to be included. |
| 2 | Batch application which produces data and ensures that the data is processed by other components and peak hours need to be given consideration. |
| 3 | Online data collection, distributed processing and data transfer in one direction. Performance to be considered critical during all business hours. |
| 4 | Tiered architecture supporting only one communications protocol with distributed processing and data transfer in both directions. Performance requirements are stringent. |
| 5 | Tiered architecture with support for several communications protocols, most appropriate component chosen dynamically for processing functions. Performance analysis done during all SDLC stages to meet user requirements. |

Table 1. DI vs. Response Time

### 7.2. Security

Security can be mapped to two GSCs – Multiple sites and Online Update. The DI mapping is detailed below.

| DI | Guidelines |
|---|---|
| 0 | One user/installation with batch processing. |
| 1 | Multiple but identical hardware and programming environments. Online update of 1-3 control files. |
| 2 | Multiple but similar hardware and programming environments. Online update of 4 or more control files. |
| 3 | Multiple sites with different hardware and programming environments. Online update of major internal logical files. |
| 4 | Application designed and supported at multiple sites for similar or identical hardware and software environments. Protection of data essential. |
| 5 | Application designed and supported for multiple sites with different hardware and programming environments. High volumes of data requiring |
| | complete data recovery procedures in place. |

Table 2. DI vs. Security

### 7.3. Availability

Availability is mapped to Online data entry and Operation Ease and the DI is as shown below.

| DI | Guidelines |
|---|---|
| 0 | One user/installation with batch processing and simple backup procedures in place. |
| 1 | Multiple identical installations with online update of 1-3 control files. Effective startup, backup and recovery procedures by operator in place. |
| 2 | Several similar installations with online update of 4 or more control files. Effective startup, backup and recovery procedures with no operator intervention. |
| 3 | Several differing sites with online update of major internal logical files. Minimal use for tape mounts. |
| 4 | Several differing sites with online update of major internal logical files. Minimal use of paper handling. Protection of data essential. |
| 5 | Several differing sites with online update of huge volumes of data requiring fully automated data recovery procedures. |

Table 3. DI vs. Availability

### 7.4. Capacity

Similar to the mapping done above Capacity can be mapped to Transaction rates and Multiple sites.

| DI | Guidelines |
|---|---|
| 0 | Single user/installation with no peaking in transactions. |
| 1 | Multiple identical installations with defined peak transaction periods like monthly, quarterly, annually, etc. |
| 2 | Multiple similar installations with weekly peaking anticipated. |
| 3 | Several differing sites with daily peak times. |
| 4 | Several differing sites with very high transaction rates requiring performance analysis included in design stage. |
| 5 | Several differing sites with very high transaction rates requiring performance analysis included in all SDLC stages. |

Table 4. DI vs. Capacity

Based on these tables the two extreme ranges of values for total degree of influence by assuming the lowest and highest degree of influence values for the seven NFRs are:

- The total degree of influence value = 0 when all the 7 NFRs have the lowest degree of influence.
- The total degree of influence value = 35 when all 7 NFRs have the highest degree of influence.

The Value Adjustment Factor (VAF) [7] can then be calculated by summing the above mentioned DI values and adding it to the TDI value obtained from the GSCs. This additional sum of DIs i.e (  DIs of NFRs ) can be called Additional DI (ADI).

$$\text{New } TDI_N = TDI + ADI, \quad (1)$$

where

$$ADI = \sum_{i=1}^{7} DI(NFR_i) \quad (2)$$

$$VAF = 0.65 + (0.01 * TDI_N) \quad (3)$$

Summing the extreme values of TDI with ADI we get

- The New $TDI_N$ value = 0 when all the 14 GSCs and 7 NFRs have the lowest degree of influence.
- The New $TDI_N$ value = 105 when all the 14 GSCs and 7 NFRs have the highest degree of influence.

If one takes the mid-range of New $TDI_N$ as average (between 0 and 105), it is obvious the New $TDI_N$ has a variation range of +52.5% to -52.5%

The Function Point Count can then be determined for development projects as

$$FP = \text{Unadjusted Function Point} * \text{Value Adjustment Factor}. \quad (4)$$

All the items mentioned above are mutually exclusive. This range can be used to modify the unadjusted Function points counted for any application to thus include an assessment of the NFRs during estimation.

This can be used to modify the estimate done using the FPA method. Further research needs to be done in this area. Data from projects need to be collected and metrics verified to determine the correctness of this approach.

## 8. Conclusion

Mapping certain NFRs and determining the degree of influence of each have been carried out in this paper. The study shows that NFRs effect the FP value and hence they also have to be accounted. Capturing the actual applicable attributes of the fourteen GSCs for a given application is very complicated. Similarly determining the influence of NFRs on the project size is also difficult. This mapping has been developed to help those in the field to include the influence of NFRs while estimating the project size and consequently the effort, schedule, priorities of tasks, etc. This hypothesis needs to be validated by measuring data from live projects and using the results to modify the mapping.